\documentclass[preprint,12pt]{elsarticle}
\usepackage{graphicx}
\usepackage{amssymb,amstext,amsmath}
\usepackage{natbib}
\usepackage{physics}
\usepackage{comment}
\newcommand{\bvarepsilon}{\boldsymbol{\varepsilon}}
\newcommand{\btau}{\boldsymbol{\tau}}
\newcommand{\bsigma}{\boldsymbol{\sigma}}
\newcommand{\bnu}{\boldsymbol{\nu}}
\newcommand{\bkappa}{\boldsymbol{\kappa}}
\newcommand{\ldbracket}{[\![}
\newcommand{\rdbracket}{]\!]}
\begin{document}
\begin{frontmatter}
\title{Safe equilibrium and crack growth in inhomogeneous materials as a variational problem}
\author{K. C. Le\footnote{corresponding author: ++49 234 32-26033, email: chau.le@rub.de}$^{a}$, M. El Yaagoubi$^b$}
\address{$^a$\,Lehrstuhl f\"ur Mechanik-Materialtheorie, Ruhr-Universit\"at Bochum, Universit\"atstr. 150, 44780 Bochum, Germany
\\
$^b$\,MS-Schramberg GmbH \& Co. KG, Max-Planck-Stra\ss e 15, 78713 Schramberg, Germany}
\begin{abstract}	
The variational principle of safe equilibrium for inhomogeneous elastic cracked bodies is formulated. Using the standard calculus of variations, we show that the crack remains in safe equilibrium as long as the maximum energy reduction rate of the virtually growing crack is negative. The crack starts to grow in the direction of the maximum energy reduction rate when the latter becomes zero. This energetic criterion implies the criteria proposed by He and Hutchinson (Int J Solids Struct 25:1053--1067, 1989). As an application we use this criterion to predict the growth direction of an interface crack in a bimaterial.
\end{abstract}
\begin{keyword}
safe equilibrium \sep crack growth \sep energy reduction rate \sep stress intensity factor \sep interface crack.
\end{keyword}

\end{frontmatter}

\section{Introduction}
The main objective of materials engineering is to improve material properties such as strength, high temperature resistance, corrosion resistance, hardness and conductivity. This goal can be achieved by joining different materials in various ways, such as adhesive bonds, protective coatings, thin film/substrate systems for electronic packages or composite bodies, to name a few. Based on their individual properties, materials such as ceramics, polymers, glasses or metals can be combined. For polycrystalline materials such as metals or alloys, which are inherently inhomogeneous, hardening improvement can be achieved by metal forming and heat treatment that change both the average grain size and the microstructure. All these inhomogeneous materials consist of at least two different homogeneous components (or grains) with the interface (or grain boundary) between them. During their manufacturing or use, various defects such as vacancies, dislocations, grain boundaries, microcracks, micropores may occur. With increasing stress level, these defects can grow and coalesce, leading to detachment, fracture or damage of the components or whole bodies. 

\begin{figure}[htp]
	\begin{tabular}{cc}
		\includegraphics[width=0.47 \textwidth]{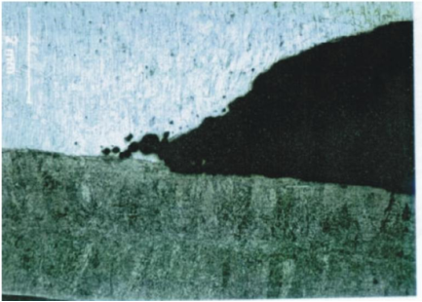} &  
		\includegraphics[width=0.48 \textwidth]{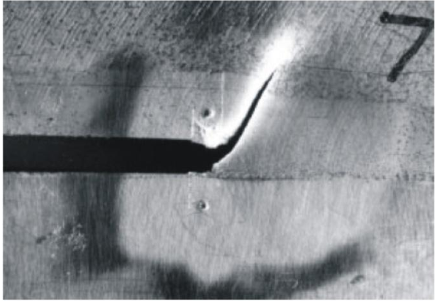} \\
		(i)  &   (ii)  
	\end{tabular}
	\caption{(Color online) Growth of an interface crack in a bimaterial: (i) Along the interface, (ii) Kinking into the softer component \cite{hao1996simulation}.}
	\label{fig:1}
\end{figure}

From the above, we immediately see the important role of fracture mechanics in predicting safe equilibrium and crack growth in these inhomogeneous materials. Fig.~\ref{fig:1} shows the competing possibilities of an interface crack in a loaded bimaterial: either it grows along the interface, or it kinks out into one of the components. To predict which of these competing possibilities will occur, we need a criterion for crack growth. However, because of the different bulk properties of the components, as well as different molecular bonding within the components and at the interface, deriving this criterion from first principles of mechanics is not as straightforward as one might think. Even the first step in solving this problem, namely the study of the stress field near the interface crack tip, encountered the logical difficulties. As first shown by Williams \cite{williams1959stresses}, the asymptotic displacement and stress fields near the crack tip exhibited oscillatory behavior. Although the zone of oscillation is estimated to be of the order of $10^{-4}$ of the crack length \cite{erdogan1963stress,england1965crack}, the oscillation of the displacement field leads to unphysical penetration of the materials (see also the discussion in \cite{malyshev1965strength}). Another difficulty associated with this oscillatory behavior is the decoupling of modes I and II in plane strain problems and the definition of the corresponding stress intensity factors \cite{rice1965plane,erdogan1971layered}. Note that when the mismatch parameter $\beta $ (introduced by Dundurs \cite{dundurs1969discussion} for bimaterials whose components are isotropically elastic) vanishes, the oscillatory behavior of the interface crack disappears and modes I and II can be uniquely separated. To get rid of the unphysical oscillatory behavior near the tip of an interface crack in the case $\beta\ne 0$, Atkinson \cite{atkinson1977stress} and Comninou \cite{comninou1977interface} each proposed a modification (see also the review article by Comninou \cite{comninou1990overview}). Atkinson acknowledged that the interface cracks are not sharp. This leads to a gradual transition and a non-oscillatory stress field. Comninou stayed with the sharp crack model, but introduced the impenetration constraint, which leads to partial closure of the crack faces and also to the elimination of stress oscillation. It is also worth noting that the oscillatory behavior does not appear in the finite deformation theory, as shown by Knowles and Sternberg for the interface crack between two neo-Hookean sheets \cite{knowles1983large}. 

However, the main unresolved issue relates to the criterion of crack growth in inhomogeneous materials and how it should be derived from first principles of mechanics. So far, a large number of quite different crack growth criteria have been proposed. They can be roughly divided into two groups: local criteria involving the stress or crack opening at the crack tip, and global criteria related to the energy release rate and the J-integral (see, e.g. \cite{richard2005theoretical} and the references therein for a review of widely used criteria). Even the most cited studies on this subject, by He and Hutchinson \cite{he1989crack,he1989kinking} and Hutchinson and Suo \cite{hutchinson1991mixed}, contain uncertainties in the choice of criteria. For the crack reaching the interface, the growth direction should be decided in two steps. First, the criterion for the crack growing along the interface is
\begin{equation}
\label{eq:1.1}
\gamma_0/\gamma _i< G_0/G_i \quad \text{for $i=1,2$,}
\end{equation}
where $\gamma_0$ and $\gamma_i$ are the surface energies (or fracture toughnesses) of the interface and components, while $G_0$ and $G_i$ are the energy release rates for the crack growing along the interface and kinking into one of the components, respectively. Although Eq.~\eqref{eq:1.1} is energetic in nature, it was not clear whether it could be derived from the variational principle of fracture mechanics. Once the criterion \eqref{eq:1.1} is violated, the next step should be to select the direction of crack kinking from either: (i) the crack will grow in the direction of mode I such that $K_{II}=0$ (local criterion) \cite{amestoy1980analytic}, (ii) the crack will grow in the direction of the maximum energy release rate (global criterion) \cite{nuismer1975energy,hayashi1981energy}. Note that the latter criterion is consistent with that derived from the variational principle of fracture mechanics for homogeneous materials \cite{stumpf1990variational,le1999determination}.

In view of these uncertainties, we aim in this work to derive the safe equilibrium and crack growth criterion from the variational principle of fracture mechanics first formulated in \cite{stumpf1990variational,le2004variational} (cf. \cite{gibbs1878art,francfort1998revisiting,bourdin2008variational,berdichevsky2009variational}). We analyze the situation when the crack tip is located on the interface, which also includes the interface crack as a special case. The crucial question is: under what condition will the crack remain in safe equilibrium and in what direction will it grow? The variational principle of fracture mechanics states that an elastic body containing a crack will remain in safe equilibrium as long as its energy functional in that state has a local minimum. By a local minimum, we mean the minimum among neighboring admissible states, including those with a virtually growing crack and with possible crack kinks. As we will see, this variational principle implies that as long as the maximum energy reduction rate of the virtually growing crack is negative, the crack remains in safe equilibrium. The crack begins to grow in the direction of the maximum energy reduction rate when the latter becomes zero. We will show that this criterion implies Eq.~\eqref{eq:1.1} in combination with the global criterion of the maximum energy release rate for the crack kinking. Since the energy release rate is expressed in terms of the stress intensity factors of the kinked crack, the problem reduces to finding the relationship between the stress intensity factors before and immediately after crack kinking. This relationship with the transformation matrix of stress intensity factors was established for homogeneous materials in \cite{amestoy1992crack} (see also \cite{wu1978elasticity,wu1978maximum}). For the interface crack, the transformation matrix of stress intensity factors can be calculated numerically by solving the singular integral equation \cite{he1989kinking} (see also \cite{he1989akinking,noijen2012semi}); this matrix shows a dependence on the kink angle and the Dundurs mismatch parameters mentioned above. Using the results obtained in \cite{he1989kinking,he1989akinking,noijen2012semi}, we can assess the safe equilibrium and predict the growth direction of the interface crack. 

This article is organized as follows. After this Introduction, we introduce in Section~2 the variational principle of safe equilibrium for inhomogeneous bodies containing cracks and derive its consequences. Section~3 deals with the calculation of the energy release rate of the interface crack as a function of the kink angle. Section~4 is devoted to evaluating the safe equilibrium and predicting the growth direction of the interface crack based on the relationship between the stress intensity factors before and immediately after crack kinking. We conclude in Section~5 with a brief summary and future research directions. 

\section{Variational principle of safe equilibrium}
\begin{figure}[htb]
\centering \includegraphics[height=6.5cm]{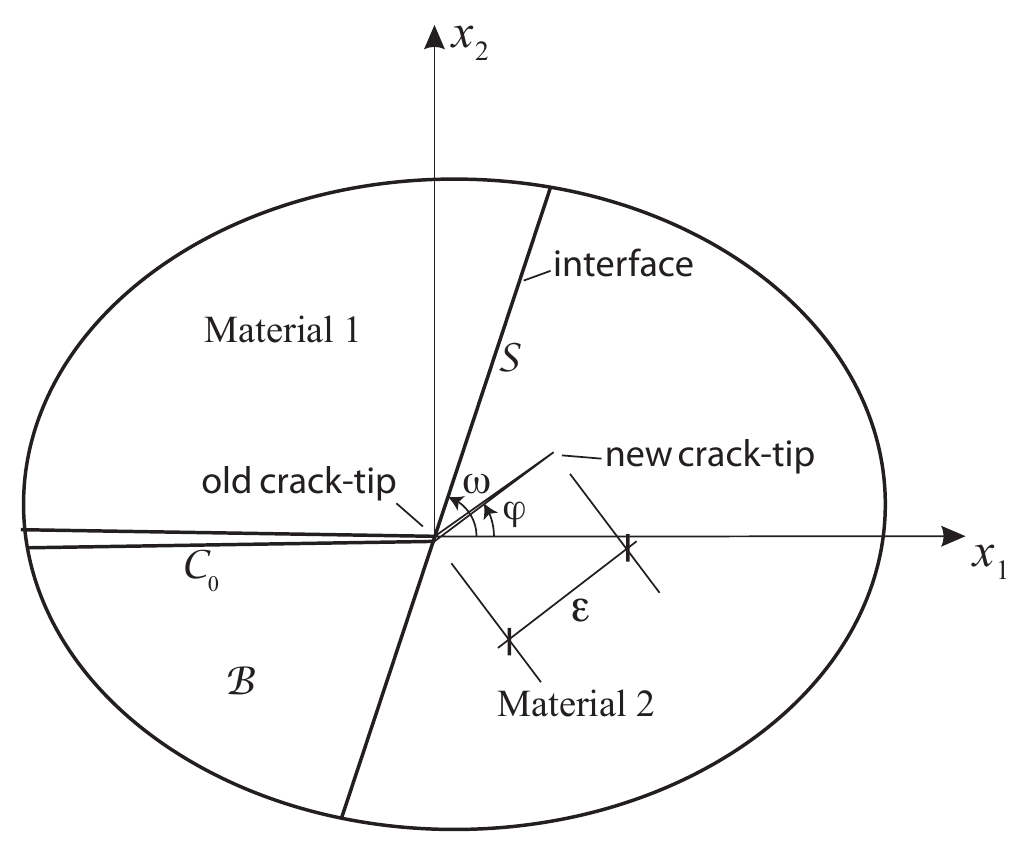} \caption{A virtually growing crack.}
\label{fig:2}
\end{figure}
For simplicity, we consider a bimaterial consisting of two different isotropic linearly elastic components that are well bonded along an interface. Let an initial configuration of this body contain a crack defined as a surface with a broken bond between adjacent material points. We restrict ourselves to the 2-D plane strain problem by considering a body of cylindrical shape. The cross section of the body occupies the region $\mathcal{B}_{\mathcal{C}_0}=\mathcal{B}\backslash \bar{\mathcal{C}_0}$ of the $(x_1,x_2)$-plane. The outer boundary
of this region, $\partial \mathcal{B}$, is decomposed into two disjoint parts $\partial _s$ and $\partial _k$, on which the tractions and displacements are given, respectively. The inner boundary is occupied by the initial crack $\bar{\mathcal{C}_0}=\mathcal{C}_0\cup \partial \mathcal{C}_0$, which, for simplicity, is assumed to have only one tip lying on the interface $\mathcal{S}$. The case of a crack with two tips or a multiple crack can be considered in a similar way. Since the bond is broken on the crack, the 2-D displacement field $\vb{u}(\vb{x})$ on $\mathcal{C}_0 \cup \partial \mathcal{C}_0$ is not defined. The limits of $\vb{u}(\vb{x})$ on two sides of the curve $\mathcal{C}_0 $ are denoted by $\vb{u}^{+}$ and $\vb{u}^{-}$. Let $\mathcal{C} \supseteq \mathcal{C}_0$ be a curve modeling a virtual growing crack and $\mathcal{B}_{\mathcal{C}}=\mathcal{B}\backslash \bar{\mathcal{C}}$. We allow for kinking of the crack, so $\mathcal{C}$ is generally assumed to be a piecewise smooth curve. There is also the possibility that the initial crack develops into multiple branches, which is very likely in the case of rapid crack propagation \cite{ravi1984bexperimental}; this situation is not considered in this paper. In the simplest case, the coordinate system is chosen so that locally $\mathcal{C}_0$ lies on the negative $x_1$-axis, while the interface $\mathcal{S}$ is the straight line inclined at an angle $\omega$ to the $x_1$-axis. Without limiting generality, we can assume that $0\le \omega<\pi$. The virtual growing crack is inclined by an angle $\varphi$ to the $x_1$-axis, as schematically shown in Fig.~\ref{fig:2}. There are five special cases: (i) $\varphi <\omega$: The crack impinges the interface, (ii) $\varphi>\omega$: The crack is reflected from the interface, (iii) $\varphi=\omega$: The crack deflects into the interface, (iv) $\omega=\varphi=0$: The interface crack continues to grow along the interface, (v) $\omega=0$, $\varphi\ne 0$: The interface crack kinks out of the interface.

The interpenetration of materials on the opposite crack faces is not allowed, hence the boundary condition
\begin{equation}
\ldbracket u_\alpha \rdbracket n_\alpha =(u^+_\alpha -u^-_\alpha )n_\alpha \ge 0
\end{equation}
must be satisfied everywhere on $\mathcal{C}$, where the Greek indices run from 1 to 2. Here $\vb{n}$ is the unit normal vector pointing in the $+$ direction. For the interface $\mathcal{S}$ between two well-bonded materials, the displacements must be continuous everywhere, unless part of $\mathcal{C}$ lies on it, and we choose the unit normal vector $\vb{n}$ on $\mathcal{S}$ to point to material 1. We define the set of kinematically admissible displacements of the body with a virtually growing crack as
\begin{equation}
\mathcal{D}=\{ \vb{u}(\vb{x})\in H^2(\mathcal{B}\backslash \mathcal{C})|\, \ldbracket u_\alpha \rdbracket n_\alpha \ge
0\, \text{on $\mathcal{C}$}, \vb{u}|_{\partial _k}=\vb{0} \} ,
\end{equation}
with $\mathcal{C}$ being an arbitrary piecewise smooth curve containing $\mathcal{C}_0$. Functions $\vb{u}(\vb{x})$ together with their first derivatives with respect to $x_\alpha $ are assumed to be square integrable in $\mathcal{B}\backslash \mathcal{C}$ and, thus, belong to the Hilbert space $H^2(\mathcal{B}\backslash \mathcal{C})$. This guarantees the finiteness of the energy per unit length in the $x_3$-direction which is defined by
\begin{equation}
\Psi [\vb{u}(\vb{x})] = \int_{\mathcal{B}_\mathcal{C}} \psi (\vb{x},\bvarepsilon(\vb{u})) \dd[2]{x} +
\int_{\mathcal{C}} \gamma (\varphi,\ldbracket \vb{u} \rdbracket ,\vb{n})\dd{s} - \int_{\partial _s}
\tau _\alpha u_\alpha \dd{s}. \label{2.4.2}
\end{equation}
In formula \eqref{2.4.2} $\psi (\vb{x},\bvarepsilon)=\frac{1}{2}\lambda (\vb{x})(\varepsilon _{\alpha \alpha })^2+\mu (\vb{x}) \varepsilon _{\alpha \beta }\varepsilon _{\alpha \beta }$ denotes the free energy density of the bimaterial, where its elastic moduli are piecewise constant functions of $\vb{x}$
\begin{equation}
\{ \lambda (\vb{x}), \mu (\vb{x})\} = \begin{cases}
 \{ \lambda _1, \mu _1 \}     & \text{for material 1}, \\
 \{ \lambda _2, \mu _2 \}       & \text{for material 2}.
\end{cases}
\end{equation}
Function $\gamma (\varphi,\ldbracket \vb{u} \rdbracket ,\vb{n})$ in the second (line) integral is the surface (cohesive) energy per unit area which may depend on the direction of crack growth $\varphi$, on the jump in displacement $\ldbracket \vb{u} \rdbracket$, and on the normal vector $\vb{n}$. Note that the crack growth in ductile materials involving the energy dissipation and the crack resistance (or fracture toughness) can also be reformulated in terms of function $\gamma$ as will be shown later. The strain tensor $\bvarepsilon$ is expressed through the displacement field by
\begin{equation}
\varepsilon _{\alpha \beta }(\vb{u}) = \frac{1}{2}(u_{\alpha ,\beta }+u_{\beta ,\alpha }). 
\end{equation}
On the part $\partial _s$ of the exterior boundary $\partial \mathcal{B}$ the traction vector $\btau$ is specified so that the last term in \eqref{2.4.2} corresponds to the work done by $\btau$. We say that the body with the pre-existing crack $\mathcal{C}_0$ is in {\it safe equilibrium} if there exist a displacement field $\check{\vb{u}}(\vb{x})$ whose discontinuity curve is $\mathcal{C}_0$ and a neighborhood $\mathcal{D}_\epsilon \subset \mathcal{D}$ of it such that the energy functional reaches a local minimum at $\check{\vb{u}}(\vb{x})$ \cite{stumpf1990variational,le2004variational}
\begin{equation}
\Psi [\check{\vb{u}}(\vb{x})]=\min_{\vb{u}(\vb{x})\in \mathcal{D}_\epsilon }\Psi[\vb{u}(\vb{x})].
\end{equation}
If this is not the case, we say that the crack begins to grow.

We are going now to establish the necessary conditions for the displacement field ${\vb{u}}(\vb{x})$ of an inhomogeneous elastic body with a pre-existing crack to be in safe equilibrium in accordance with the principle of minimum energy given above. Following Griffith \cite{griffith1920phenomena}, we assume that the surface energy does not depend on $\ldbracket \vb{u} \rdbracket $ and $\vb{n}$, so $\gamma (\varphi,\ldbracket \vb{u} \rdbracket ,\vb{n})=\gamma (\varphi)$, where $\gamma(\varphi)=\gamma_1$ if the crack is reflected into material 1, $\gamma(\varphi)=\gamma_2$ if the crack impinges the interface and grows in material 2, and $\gamma(\varphi)=\gamma_0$ if the crack deflects into the interface. Let us introduce a one parameter family of admissible displacements $\epsilon\mapsto \vb{u}(.,\epsilon ) \in \mathcal{D}$, whose discontinuity curves describe a virtually growing crack $\mathcal{C} _{\epsilon }$ (see Fig.~\ref{fig:2}). Since the crack can only grow, we require that
\begin{gather} 
\mathcal{C} _{\epsilon '}\supseteq \mathcal{C} _{\epsilon }\supseteq  \mathcal{C}_0 \quad
\text{for $\epsilon '>\epsilon >0$}, \quad \mathcal{C} _{\epsilon
}=\mathcal{C}_0 \quad \text{when $\epsilon =0$},
\\
\vb{u}(\vb{x},0) = \check{\vb{u}}(\vb{x}). \notag
\end{gather}
After substituting this one parameter family of admissible displacements into the energy functional \eqref{2.4.2} it becomes a function of $\epsilon $, $\epsilon \ge 0$. If the body is in safe equilibrium, this function has an end-point minimum at $\epsilon =0$ for arbitrary families of admissible displacements in accordance with our variational principle. Therefore the following necessary condition for safe equilibrium must be fulfilled:
\begin{equation}
\delta \Psi =\lim_{\epsilon \to 0}\dv{\epsilon }
\Psi [\vb{u}(.,\epsilon )] \ge 0 \quad \text{for arbitrary families of
$\vb{u}(.,\epsilon )\in \mathcal{D}$}. \label{2.4.5}
\end{equation}

\begin{figure}[htb]
\centering \includegraphics[height=4.5cm]{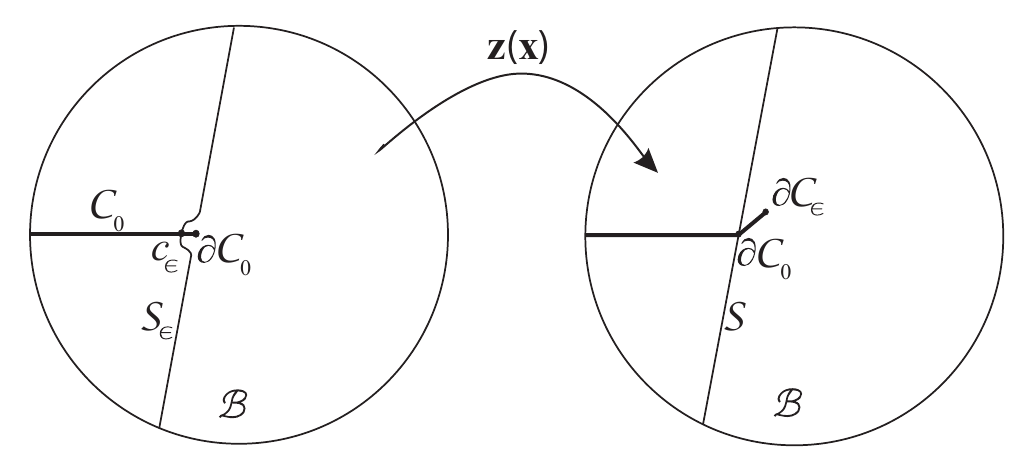} \caption{Parametrization and
integration domains.} \label{fig:3}
\end{figure}

In order to derive consequences from \eqref{2.4.5}, we must be able to calculate the derivative of $\Psi [\vb{u}(.,\epsilon )]$ with respect to $\epsilon $ and then take the limit as $\epsilon \to 0$. The difficulty of this calculation is due to the changeable region $\mathcal{B}\backslash \mathcal{C} _\epsilon $ and curve $\mathcal{C}_{\epsilon }$. In order to overcome it we introduce a one-parameter family of one-to-one mappings of $\mathcal{B}$ onto itself
\begin{equation}
\epsilon \mapsto \vb{z}(\vb{x},\epsilon )
\end{equation}
so that
\begin{gather} \label{mappings}
\mathcal{B}\backslash \mathcal{C}_0 \overset{\vb{z}}{\mapsto }
\mathcal{B}\backslash \mathcal{C}_\epsilon ,\quad \mathcal{C}_0
\overset{\vb{z}}{\mapsto } \mathcal{C}_\epsilon ,
\\
\vb{z}(\vb{x},\epsilon ) = \vb{x} , \quad \text{when $\epsilon =0$ or $\vb{x}\in
\partial \mathcal{B}$}. \notag
\end{gather}
Since the crack kinking is admitted, functions $\vb{z}(\vb{x},\epsilon )$ are assumed to be smooth everywhere except at the points $\vb{c}_\epsilon $ lying on the $x_1$-axis which are mapped to the old crack tip $\partial \mathcal{C}_0$. Noticeable are also the curves $\mathcal{S}_\epsilon$ going through $\vb{c}_\epsilon $ that are mapped to the real interface $\mathcal{S}$ (see Fig.~\ref{fig:3}). We choose $\vb{z}(\vb{x},\epsilon )$ such that $\mathcal{S}_\epsilon$ coincides with $\mathcal{S}$ except a small neighborhood of $\partial \mathcal{C}_0$. We call $\vb{z}(\vb{x},\epsilon )$ parametrizations of medium. These mappings will be used as changes of variables for the 2-D and 1-D integrals in \eqref{2.4.2} which become then integrals over the fixed region $\mathcal{B}\backslash \mathcal{C}_0$ and curve $\mathcal{C}_0$ (see Eq.~\eqref{mappings}). According to the transformation rule we have
\begin{equation}
\int_{\mathcal{B}\backslash \mathcal{C}_\epsilon } \psi (\vb{x},\bvarepsilon (\vb{x}))\dd[2]{x}  = \int_{\mathcal{B}\backslash \mathcal{C}_0}
\psi (\vb{z}(\vb{x},\epsilon ),\bvarepsilon(\vb{z}(\vb{x},\epsilon )))J\dd[2]{x},
\end{equation}
where $J=\det z_{\alpha ,\beta }$ is the Jacobian of transformation. Since the region of integration, after this change of variables, does not depend on $\epsilon $, the order of differentiation and integration can be interchanged so that
\begin{equation}
\lim_{\epsilon \to 0}
 \dv{\epsilon } \int_{\mathcal{B}\backslash \mathcal{C}_\epsilon }
\psi (\vb{x},\bvarepsilon)\dd[2]{x}=\lim_{\epsilon \to 0}
\int_{\mathcal{B}\backslash \mathcal{C}_0} (J\delta \psi + \psi \delta
J)\dd[2]{x}.
\end{equation}
Symbol $\delta $ under integral signs, called for short variation, is used to denote the partial derivative with respect to $\epsilon $ at fixed $\vb{x}$. The variation of $J$ reads
\begin{equation}
\delta J=\frac{\partial J}{\partial z_{\alpha ,\beta }}\delta
z_{\alpha ,\beta }=(\text{Cof})_{\alpha \beta }\delta z_{\alpha
,\beta }= J\frac{\partial x_\beta }{\partial z_\alpha }\delta
z_{\alpha ,\beta }=J\frac{\partial \delta z_\alpha }{\partial
z_\alpha },
\end{equation}
where $(\text{Cof})_{\alpha \beta }$ is the cofactor of $z_{\alpha ,\beta }$. Consider now the variation of $\psi $
\begin{equation}
\delta \psi =\pdv{\psi}{z_\alpha}\delta z_\alpha+\frac{\partial \psi }{\partial \varepsilon _{\alpha \beta }}\, \delta
\varepsilon _{\alpha \beta }=\pdv{\psi}{z_\alpha}\delta z_\alpha+\frac{1}{2}\sigma _{\alpha \beta }\, \delta
(\frac{\partial u_\alpha }{\partial z_\beta }+ \frac{\partial
u_\beta }{\partial z_\alpha })=\pdv{\psi}{z_\alpha}\delta z_\alpha+\sigma _{\alpha \beta }\, \delta
\frac{\partial u_\alpha }{\partial z_\beta }.
\end{equation}
Here $\bsigma$ is the symmetric stress tensor field with components $\sigma _{\alpha \beta } =\frac{\partial \psi }{\partial \varepsilon _{\alpha \beta }}$. In order to calculate the variation $\delta \frac{\partial u_\alpha }{\partial z_\beta }$ we recall the following identity
\begin{equation}
\frac{\partial u_\alpha }{\partial x_\gamma }=\frac{\partial
u_\alpha }{\partial z_\beta }\frac{\partial z_\beta }{\partial
x_\gamma }.
\end{equation}
Applying the product rule of differentiation to this identity and remembering that the variation and partial derivatives with respect to $x_\alpha $ are commutative we obtain
\begin{equation}
\frac{\partial \delta u_\alpha }{\partial x_\gamma }=\delta
\frac{\partial u_\alpha }{\partial z_\beta }\frac{\partial z_\beta
}{\partial x_\gamma }+\frac{\partial u_\alpha }{\partial z_\beta
}\frac{\partial \delta z_\beta }{\partial x_\gamma }.
\end{equation}
Multiplying this equation with $\partial x_\gamma /\partial z_\kappa $ and rearranging indices and terms we get
\begin{equation}
\delta \frac{\partial u_\alpha }{\partial z_\beta }=\frac{\partial
\delta u_\alpha }{\partial z_\beta }-\frac{\partial u_\alpha
}{\partial z_\gamma }\frac{\partial \delta z_\gamma }{\partial
z_\beta }.
\end{equation}

\begin{figure}[htb]
\centering \includegraphics[height=4.5cm]{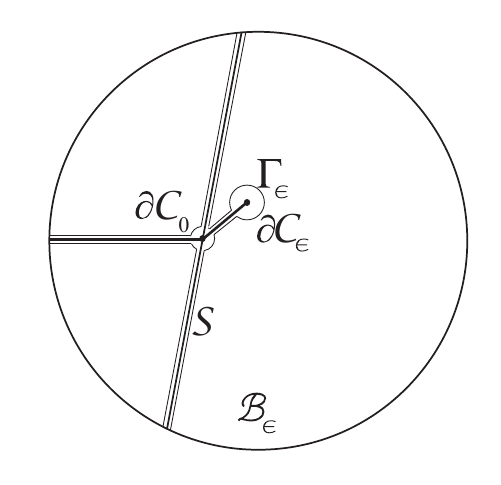} \caption{Regularized
integration domain $\mathcal{B}_\epsilon $.} \label{fig:4}
\end{figure}

Combining all these formulas we obtain finally
\begin{multline}\label{2.4.6}
\lim_{\epsilon \to 0} \dv{\epsilon }
\int_{\mathcal{B}\backslash \mathcal{C}_\epsilon } \psi (\vb{x},\bvarepsilon (\vb{u}(\vb{x},\epsilon
)) \dd[2]{x} = \lim_{\epsilon \to 0}\int_{\mathcal{B}\backslash 
\mathcal{C}_0}\Bigl( \sigma _{\alpha \beta }\frac{\partial \delta
u_\alpha }{\partial z_\beta } + \pdv{\psi}{z_\alpha}\delta z_\alpha+\mu _{\alpha \beta }\frac{\partial
\delta z_\alpha }{\partial z_\beta }\Bigr) J\dd[2]{x}
\\
=\lim_{\epsilon \to 0}\int_{\mathcal{B}\backslash 
\mathcal{C}_\epsilon }(\sigma _{\alpha \beta }\delta u_{\alpha
,\beta } + \pdv{\psi}{x_\alpha}\delta z_\alpha+ \mu _{\alpha \beta }\delta z_{\alpha ,\beta })\dd[2]{x},
\end{multline}
where 
\begin{equation}
\mu _{\alpha \beta } = - \sigma _{\gamma \beta }u_{\gamma ,\alpha
} + \psi \delta _{\alpha \beta }
\end{equation}
is the Eshelby tensor \cite{eshelby1951force}, while $\pdv*{\psi}{x_\alpha}$ denotes the partial derivatives of $\psi$ at fixed $\bvarepsilon$. For the piecewise constant functions $\lambda (\vb{x})$ and $\mu (\vb{x})$ the derivative $\pdv*{\psi}{x_\alpha}$ must be understood as the generalized function. Since the stress field $\bsigma$ and the displacement gradients $u_{\alpha ,\beta }$ are singular at $\partial \mathcal{C}_0$ and $\partial \mathcal{C}_\epsilon $ and may suffer jumps on $\mathcal{S}$, Gauss' theorem cannot be applied to the right-hand side of \eqref{2.4.6} directly. To do this properly we replace the region $\mathcal{B}\backslash \mathcal{S}_\epsilon $ by $\mathcal{B}_\epsilon $, whose interior boundary is shown in Fig.~\ref{fig:4}. It turns out that formula \eqref{2.4.6} remains valid for the integral taken over $\mathcal{B}_\epsilon $. Applying Gauss' theorem and letting $\epsilon $ approach zero, we obtain
\begin{multline}
\lim_{\epsilon \to 0} \dv{\epsilon }
\int_{\mathcal{B}\backslash \mathcal{C}_\epsilon }\psi (\vb{x},\bvarepsilon(\vb{u}))\dd[2]{x} =
\int_{\mathcal{B}\backslash \mathcal{C}_0} [-{\sigma
}_{\alpha \beta ,\beta }\delta u_\alpha +(\pdv{\psi}{x_\alpha}- {\mu }_{\alpha
\beta ,\beta })\delta z_\alpha ]\dd[2]{x} 
\\
+ \int_{\mathcal{C}_0}[(-{\sigma }^{+}_{\alpha \beta } \delta
u^{+}_{\alpha }
+ {\sigma }^{-}_{\alpha \beta }\delta u^{-}_{\alpha
})n_{\beta } + (-{\mu }^{+}_{\alpha \beta }+ {\mu
}^{-}_{\alpha \beta })n_{\beta } \delta z_{\alpha }]\dd{s}
\\
+ \int_{\mathcal{S}}(-{\sigma }^{+}_{\alpha \beta } 
+ {\sigma }^{-}_{\alpha \beta })\delta u_{\alpha
}n_{\beta } \dd{s}+\int_{\partial _s}{\sigma }_{\alpha \beta }\delta
u_\alpha n_\beta \dd{s} - G(\varphi )\delta l . \label{2.4.7}
\end{multline}
In the last term of \eqref{2.4.7} $G(\varphi )$ is given by the J-integral \cite{rice1968path}
\begin{equation}
G(\varphi )=\nu _\alpha \lim_{\epsilon \to 0}\int_{\Gamma _\epsilon }
\mu _{\alpha \beta }\kappa _{\beta }\dd{s} = \nu _\alpha
\lim_{\epsilon \to 0} \int_{\Gamma _\epsilon }(-\sigma _{\gamma
\beta }u_{\gamma ,\alpha }\kappa _{\beta }+ \psi \kappa _{\alpha })
\dd{s} , \label{2.4.8}
\end{equation}
with $\bnu$ being the unit vector pointing to the direction of crack extension, $\Gamma _\epsilon $ the contour of radius $\epsilon _1\ll \epsilon$ surrounding the crack tip $\partial \mathcal{C}_\epsilon $, $\bkappa$ the outward unit normal vector on $\Gamma _\epsilon $, and $\delta l=\nu _\alpha \lim_{\epsilon \to 0}\delta z_\alpha |_{\partial \mathcal{C}_\epsilon }$ the virtual crack extension length. Note that, away from the crack tip, the stress field $\bsigma$, the displacement gradients $u_{\alpha ,\beta }$ and the energy density $\psi $ approach the corresponding quantities calculated for $\vb{u}=\check{\vb{u}}$ in the limit $\epsilon \to 0$. We drop for short the check over these quantities. Note also that the integral along the contour surrounding the point $\partial \mathcal{C}_0$ tends to zero as $\epsilon \to 0$, because the stress field and the displacement gradients have in its neighborhood the corner singularity, which turns out to be weaker than the square root singularity at the crack tip \cite{hein1971stress}. When deriving \eqref{2.4.7} the following asymptotic property is tacitly used
\begin{equation}
\lim_{\epsilon \rightarrow 0} \int_{\Gamma _\epsilon } \sigma
_{\alpha \beta }\kappa _\beta \dd{s} = 0.
\end{equation}
This is due to the square root singularity of the stress field \cite{williams1959stresses}. The integral \eqref{2.4.8} describes the elastic energy release per unit crack extension length which is expected to be a function of the kink angle $\varphi $.

The variation of the surface energy is given by
\begin{equation}
\lim_{\epsilon \to 0} \dv{\epsilon } \int_{\mathcal{C}
_\epsilon } \gamma (\varphi) \dd{s} =\gamma (\varphi) \delta l. 
\end{equation}
Note that, if there is resistance to crack growth and, consequently, nonzero energy dissipation, the variational inequality \eqref{2.4.5} must be extended as follows
\begin{equation}
\delta \Psi +r(\varphi) \delta l \ge 0 \quad \text{for arbitrary families of
$\vb{u}(.,\epsilon )\in \mathcal{D}$}, \label{2.4.5a}
\end{equation}
where $r(\varphi)$ is the fracture toughness. However, if the latter does not depend on the crack tip velocity (rate-independent theory), the second term on the left-hand side of \eqref{2.4.5a} can be written as
\begin{equation}
r(\varphi) \delta l =\lim_{\epsilon \to 0} \dv{\epsilon } \int_{\mathcal{C}
_\epsilon } r(\varphi) \dd{s},
\end{equation}
so it can be combined with the variation of the surface energy term, with $\gamma_\text{eff}=\gamma +r$ being interpreted as the effective surface energy density. In this sense, crack growth in ductile materials involving energy dissipation and rate-independent fracture toughness can also be reformulated in terms of $\gamma_\text{eff}$. The variation of the external work is equal to
\begin{equation}
\lim_{\epsilon \to 0} \dv{\epsilon } \int_{\partial _s}
\tau _\alpha u_\alpha \dd{s} = \int_{\partial _s} \tau _\alpha
\delta u_\alpha \dd{s}. \label{2.4.10}
\end{equation}

Combining formulas \eqref{2.4.7}-\eqref{2.4.10}, one transforms the inequality \eqref{2.4.5} to
\begin{multline}
\int_{\mathcal{B}\backslash \mathcal{C}_0} [-{\sigma }
_{\alpha \beta ,\beta }\delta u_\alpha +(\pdv{\psi}{x_\alpha}- {\mu }_{\alpha
\beta ,\beta })\delta z_\alpha ]\dd[2]{x} + \int_{\mathcal{C}_0}[
(-{\sigma }^{+}_{\alpha \beta } \delta u^{+}_{\alpha }+
{\sigma }^{-}_{\alpha \beta }\delta u^{-}_{\alpha })n_{\beta
}
\\
+ (-{\mu }^{+}_{\alpha \beta }+ {\mu }^{-}_{\alpha
\beta })n_{\beta } \delta z_{\alpha }] \dd{s} + \int_{\mathcal{S}}(-{\sigma }^{+}_{\alpha \beta } 
+ {\sigma }^{-}_{\alpha \beta })\delta u_{\alpha
}n_{\beta } \dd{s}
\\
+\int_{\partial _s}({\sigma }_{\alpha \beta }n_\beta -\tau _\alpha )\delta
u_\alpha \dd{s} - [G(\varphi )-\gamma (\varphi) ]\delta l \ge 0. \label{2.4.11}
\end{multline}

We shall now analyze the inequality \eqref{2.4.11}. It is obvious that the variations $\delta \vb{u} $ and $\delta \vb{z}$ in the region $\mathcal{B}\backslash \mathcal{C}_0$ as well as $\delta \vb{u}$ on $\partial _\tau $ can be chosen arbitrarily. On the contrary, $\delta \vb{u}^\pm $ and $\delta \vb{z}$ on $\mathcal{C}_0$ should satisfy some constraints. When the crack faces are not in contact with each other, then the variations $\delta \vb{u}^{\pm }$ on $\mathcal{C}_0$ can obviously have arbitrary values. If this is not the case, the constraint $\ldbracket \delta u_\alpha \rdbracket n_\alpha \ge 0$ must be obeyed. Since $\mathcal{C} _{\epsilon } \supseteq \mathcal{C}_0$, $\delta z_\alpha$ should satisfy the constraints
\begin{equation}
\delta z_\alpha n_\alpha = 0 \quad \text{on $\mathcal{C}_0$},\quad
\delta l\ge 0\quad \text{at $\partial \mathcal{C}_0$}.
\end{equation}

Taking all these constraints into account, one can show that the variational inequality \eqref{2.4.11} leads to
\begin{gather}
{\sigma }_{\alpha \beta ,\beta } = 0 ,\quad {\sigma }
_{\alpha \beta } = \frac{\partial \psi } {\partial {\varepsilon } _{\alpha
\beta }} \quad \text{in} \quad \mathcal{B}\backslash \mathcal{C}_0
, \notag
\\
{u}_\alpha = 0 \quad \text{on $\partial _k$} , \quad
{\sigma }_{\alpha \beta } n_\beta = \tau _\alpha \quad
\text{on $\partial _s$}, \notag
\\
{\sigma }^{+}_{\alpha \beta } n_{\beta }=
{\sigma }^{-}_{\alpha \beta } n_{\beta } \quad
\text{on $\mathcal{S}$}, \notag
\\
({u}_\alpha ^+-{u}_\alpha ^-)n_\alpha \ge 0\quad
\text{on $\mathcal{C}_0$,} \label{2.4.12}
\\
{\sigma }^{+}_{\alpha \beta }n_\beta ={\sigma }
^{-}_{\alpha \beta }n_\beta = -p n_\alpha ,\quad p\ge 0 \quad
\text{on $\mathcal{C}_0$}, \notag
\\
({u}_\alpha ^+-{u}_\alpha ^-)n_\alpha > 0 \Rightarrow
p=0, \notag
\\
\max_{\varphi }[G(\varphi )-\gamma (\varphi)]\le  0 \quad \text{at $\partial
\mathcal{C}_0$}. \notag
\end{gather}
Additionally, we also get the relations
\begin{gather}
\pdv{\psi}{x_\alpha}-{\mu }_{\alpha \beta ,\beta } = 0 ,\quad {\mu
}_{\alpha \beta } = - {\sigma }_{\gamma \beta }
{u}_{\gamma ,\alpha} + {\psi }\delta _{\alpha \beta }
\quad \text{in} \quad \mathcal{B}\backslash \mathcal{C}_0 ,
\\
(-{\mu }_{\alpha \beta }^++{\mu }_{\alpha \beta
}^-)n_\beta \nu _\alpha =0 \quad \text{on $\mathcal{C}_0$}, \notag
\end{gather}
with $\bnu$ being the tangent vector to the curve $\mathcal{C}_0$. However, it is easy to see that these equations are satisfied identically by virtue of other equations in \eqref{2.4.12}. This is due to the invariant properties of the energy functional with respect to the group of parametrizations leaving the curve of discontinuity unchanged.

Thus, equations \eqref{2.4.12} are the necessary conditions for the displacement field of the cracked body $\check{\vb{u}}$ to be in safe equilibrium. The difference between equilibrium and safe equilibrium reduces to the last condition, $\max_{\varphi }[G(\varphi )-\gamma (\varphi)]\le 0$, called the maximum energy reduction rate criterion. Thus, if this maximum is negative, the crack stays in stable safe equilibrium. The crack starts to grow in the direction $\check{\varphi}$ that maximizes $G(\varphi )-\gamma (\varphi)$ if the maximum is equal to zero. We want now to show that this criterion implies condition \eqref{eq:1.1} for the crack growing along the interface combined with the maximum energy release rate criterion for the crack kinking out into one of the components. Indeed, according to the maximum energy reduction rate criterion, the crack will grow along the interface if the maximum of $G(\varphi )-\gamma (\varphi)$, achieved at $\varphi=\omega$, is equal to zero. This means
\begin{equation}
0=G(\omega)-\gamma_0>G(\varphi)-\gamma(\varphi) \quad \text{for all $\varphi\ne \omega$}.
\end{equation}
Bringing the term $-\gamma(\varphi)$ to the left-hand side of this inequality and dividing by the positive constant $\gamma_0=G(\omega)$, we obtain
\begin{equation}
\label{2.14}
\gamma(\varphi)/\gamma_0 > G(\varphi)/G(\omega) \quad \text{for all $\varphi\ne \omega$},
\end{equation}
which is equivalent to \eqref{eq:1.1}. In case the condition \eqref{2.14} is not fulfilled, the maximum must be sought in one of the components ($\varphi>\omega$ or $\varphi<\omega$). Since $\gamma(\varphi)$ is constant there, the maxima of the energy reduction rate $G(\varphi )-\gamma (\varphi)$ and of the energy release rate $G(\varphi )$ are achieved at the same angle $\check{\varphi}$. Thus, the maximum energy reduction rate criterion implies the maximum energy release rate criterion in this case. Thus, to assess the safe equilibrium and predict the direction of crack growth we must find the relationship between $G(\varphi)$ and $G(\omega)$. This task will be done for the special case of the interface crack with $\omega=0$ in the next two Sections.

\section{Energy release rate of the interface crack}

\begin{figure}[htp]
	\begin{tabular}{cc}
		\includegraphics[width=0.49 \textwidth]{{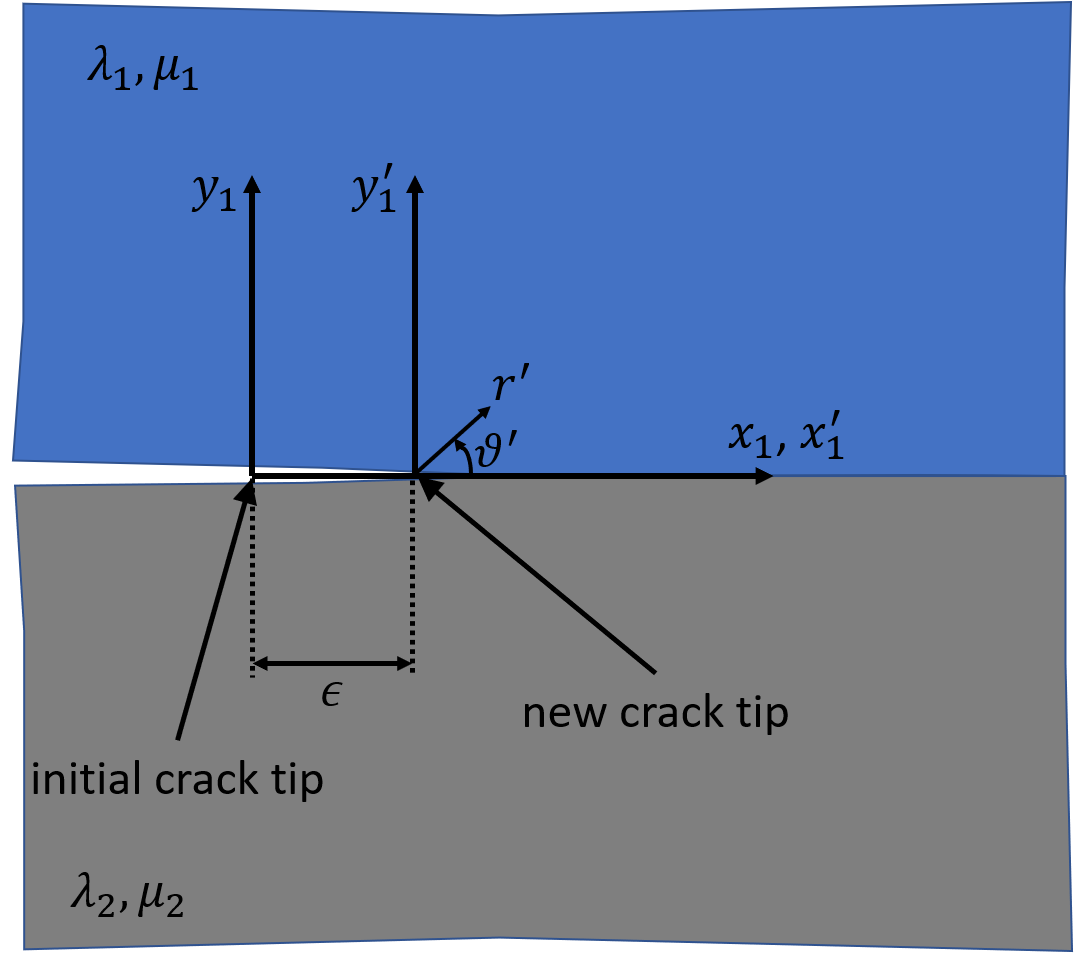}} &  
		\includegraphics[width=0.49 \textwidth]{{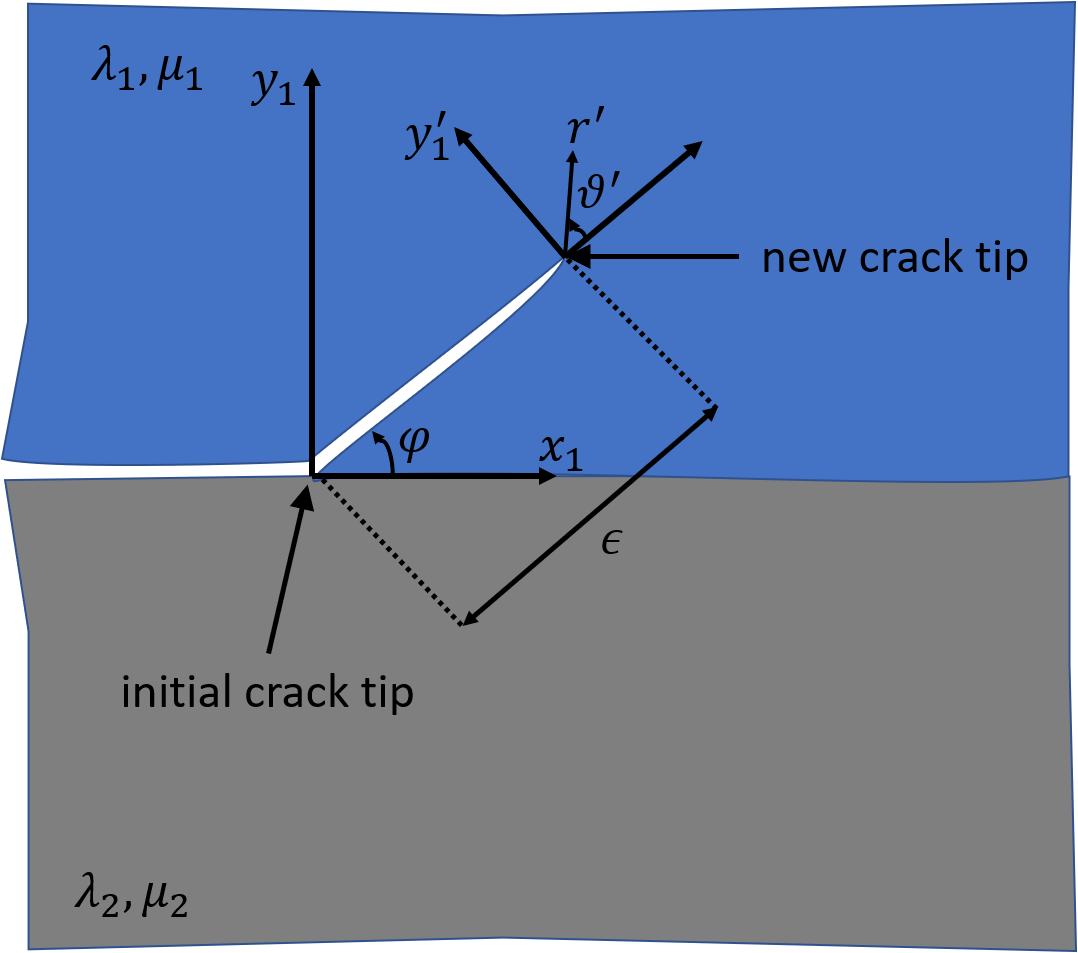}} \\
		(i)  &   (ii)  
	\end{tabular}
	\caption{(Color online) Growth of an interface crack in a bimaterial: (i) Along the interface, (ii) Kinking-out into one of the components.}
	\label{fig:5}
\end{figure}

We now apply the developed theory to an interface crack ($\omega=0$). As shown in Fig.~\ref{fig:5}, the virtual crack growth can be along the interface ($\varphi=0$), kinking out into material 1 ($\varphi>0$) or into material 2 ($\varphi<0$). We will calculate the energy release rate $G(\varphi)$, given by Eq.~\eqref{2.4.8}, in these cases. To focus on the crack growth criterion, we assume for simplicity that one of the Dundurs parameters, $\beta$, vanishes
\begin{equation}
\beta = \frac{(1-2\nu_2)/\mu_2-(1-2\nu_1)/\mu_1}{2[(1-\nu_2)/\mu_2+(1-\nu_1)/\mu_1]}=0,
\end{equation}
where $\nu_1$ and $\nu_2$ are Poisson's ratios of the corresponding components. In contrary, the other Dundurs parameter
\begin{equation}
\alpha =\frac{(1-\nu_2)/\mu_2-(1-\nu_1)/\mu_1}{(1-\nu_2)/\mu_2+(1-\nu_1)/\mu_1}
\end{equation}
is not equal to zero. The general case with $\alpha\ne 0$ and $\beta\ne 0$ is considered in \cite{he1989kinking}.

For the interface crack growing along the interface with $\varphi=0$ and $\bnu =(1,0)$, we have
\begin{equation}
\label{3.1}
G(0)=\lim_{\epsilon \to 0} \int_{\Gamma _\epsilon }(-\sigma _{\gamma
\beta }u_{\gamma ,1}\kappa _{\beta }+ \psi \kappa _1)\dd{s}.
\end{equation}
One can substitute Williams' stress and displacement fields around the crack tip into \eqref{3.1} to compute $G(0)$. However, the shorter and more convenient way is that originally proposed by Irwin \cite{irwin1957analysis}. Namely,
\begin{equation}
G(0)=\frac{1}{2\epsilon}\int_0^\epsilon \sigma^{(1)}_{\alpha \beta}(r) n_\beta \ldbracket u^{(2)}_\alpha (\epsilon -r)\rdbracket \dd{r},
\end{equation}
where $\sigma^{(1)}_{\alpha \beta}(r)$ are the stresses ahead the crack tip (at $\theta=0$) {\it prior} to the growth, while  $\ldbracket u^{(2)}_\alpha (r') \rdbracket$ are the jump in displacement behind the crack tip (at $\theta '=\pm \pi$, $r'=\epsilon-r$) {\it just after} the crack advances. Using the stress distribution ahead the initial crack 
\begin{equation}
\sigma^{(1)}_{22}(r)=K_1\frac{1}{\sqrt{2\pi r}}, \quad \sigma^{(1)}_{12}(r)=K_2\frac{1}{\sqrt{2\pi r}},
\end{equation}
and the jump in displacements behind the advanced crack
\begin{equation}
\ldbracket u^{(2)}_{2}(\epsilon-r)\rdbracket =\frac{8}{E_*}K_1\sqrt{\frac{\epsilon-r}{2\pi }}, \quad \ldbracket u^{(2)}_{1}(\epsilon-r)\rdbracket =\frac{8}{E_*}K_2\sqrt{\frac{\epsilon-r}{2\pi }},
\end{equation}
with $E_*$ being defined as
\begin{equation}
\frac{1}{E_*}= \frac{1}{4}\Bigl( \frac{1-\nu_1}{\mu_1}+\frac{1-\nu_2}{\mu_2}\Bigr),
\end{equation}
we find that \cite{malyshev1965strength}
\begin{equation}
G(0)=\frac{1}{E_*}(K_1^2+K_2^2).
\end{equation}

For the crack kinking out into material 1 ($\varphi>0$), we have
\begin{equation}
\label{3.6}
G(\varphi )=\lim_{\epsilon \to 0}\int_{\Gamma _\epsilon }(-\sigma ^\prime
_{\alpha ^\prime \beta ^\prime }u^\prime _{\alpha ^\prime
,1^\prime }\kappa _{\beta ^\prime }+ \psi ^\prime \kappa _{1^\prime })\dd{s}, 
\end{equation}
where $x^\prime _\alpha$ are the shifted and rotated coordinate system shown in Fig.~\ref{fig:5}(ii). We substitute the asymptotic formulas of the stress and displacement fields near the extended crack tip $\partial \mathcal{C}_\epsilon $ given by
\begin{equation}
\begin{split}
\sigma ^\prime _{\alpha ^\prime \beta ^\prime }(\epsilon )=
K^\prime _{I}(\epsilon ) \frac{f_{I\alpha ^\prime \beta ^\prime
}(\vartheta ^\prime )}{\sqrt{2\pi r^\prime }}+K^\prime _{II}(\epsilon
) \frac{f_{II\alpha ^\prime \beta ^\prime }(\vartheta ^\prime
)}{\sqrt{2\pi r^\prime }}+O(1),
\\
u^\prime _{\alpha ^\prime }(\epsilon )=K^\prime _{I}(\epsilon )
\sqrt{\frac{r^\prime } {2\pi }} v_{I\alpha ^\prime }(\vartheta
^\prime )+K^\prime _{II}(\epsilon )\sqrt{\frac{r^\prime }{2\pi }}
v_{II\alpha ^\prime }(\vartheta ^\prime )+O(r^\prime ),
\end{split}
\end{equation}
into \eqref{3.6}, with $r^\prime $ and $\theta^\prime $ the corresponding polar coordinates and $f_{I\alpha ^\prime \beta ^\prime }(\vartheta ^\prime )$, $f_{II\alpha ^\prime \beta ^\prime
}(\vartheta ^\prime )$, $v_{I\alpha ^\prime }(\vartheta ^\prime )$, $v_{II\alpha ^\prime }(\vartheta
^\prime )$ the well-known angular distributions of the stress and displacement fields (see, e.g., \cite{le2010introduction}). Computing the integral and letting $\epsilon$ go to zero, we get
\begin{equation}
G(\varphi )=\frac{1-\nu_1}{2\mu_1}(K_{I}^2+K_{II}^2),
\end{equation}
where $K_{I}$ and $K_{II}$ are the limiting values of the stress intensity factors $K'_{I}(\epsilon)$ and $K'_{II}(\epsilon)$ of the kinked crack when $\epsilon$ goes to zero. Similarly, for the crack kinking out into material 2 ($\varphi<0$),
\begin{equation}
\label{3.8}
G(\varphi )=\frac{1-\nu_2}{2\mu_2}(K_{I}^2+K_{II}^2).
\end{equation}
Thus, the computation of $G(\varphi)/G(0)$ reduces to finding the relationship between $(K_I,K_{II})$ and $(K_1,K_2)$. This will be done in the next Section.

\section{Growth of the interface crack}

Let us assume for definiteness that 
\begin{equation}
\max_{\varphi >0}[G(\varphi)-\gamma_1]< \max_{\varphi <0}[G(\varphi)-\gamma_2].
\end{equation}
In this case, the kinking of the crack in material 1 is excluded, and we need just to find $G(\varphi)$ for negative $\varphi$. Let us redefine $\varphi$ as the positive kink angle when the crack kinks out into material 2. The relationship between $(K_I,K_{II})$ and $(K_1,K_2)$ in the complex form reads \cite{he1989kinking}
\begin{equation}
\label{4.2}
K_I+iK_{II}=c(\varphi,\alpha)K+\bar{d}(\varphi,\alpha)\bar{K},
\end{equation}
where the bar indicates complex conjugation, $K=K_1+iK_2$, and $c$ and $d$ are complex-valued functions of $\varphi $. Substitution of \eqref{4.2} into \eqref{3.8} yields
\begin{equation}
G(\varphi )=\frac{1-\nu_2}{2\mu_2}[(\abs{c}^2+\abs{d}^2)K\bar{K}+2\Re (cdK^2)].
\end{equation}
Using the representation $K=\abs{K}e^{i \gamma}$, with $\gamma=\arctan (K_2/K_1)$ being the parameter characterizing the load combination, we bring the energy release rate ratio to the following form
\begin{equation}
\frac{G(\varphi)}{G(0)}=(1+\alpha)[(\abs{c}^2+\abs{d}^2)+2\Re (cd e^{i \gamma})]
\end{equation}
Note that, except $\varphi$, this ratio depends also on $\alpha$ and $\gamma$ through the coefficients $c$ and $d$.

\begin{figure}[htp]
	\begin{tabular}{cc}
		\includegraphics[width=0.49\textwidth]{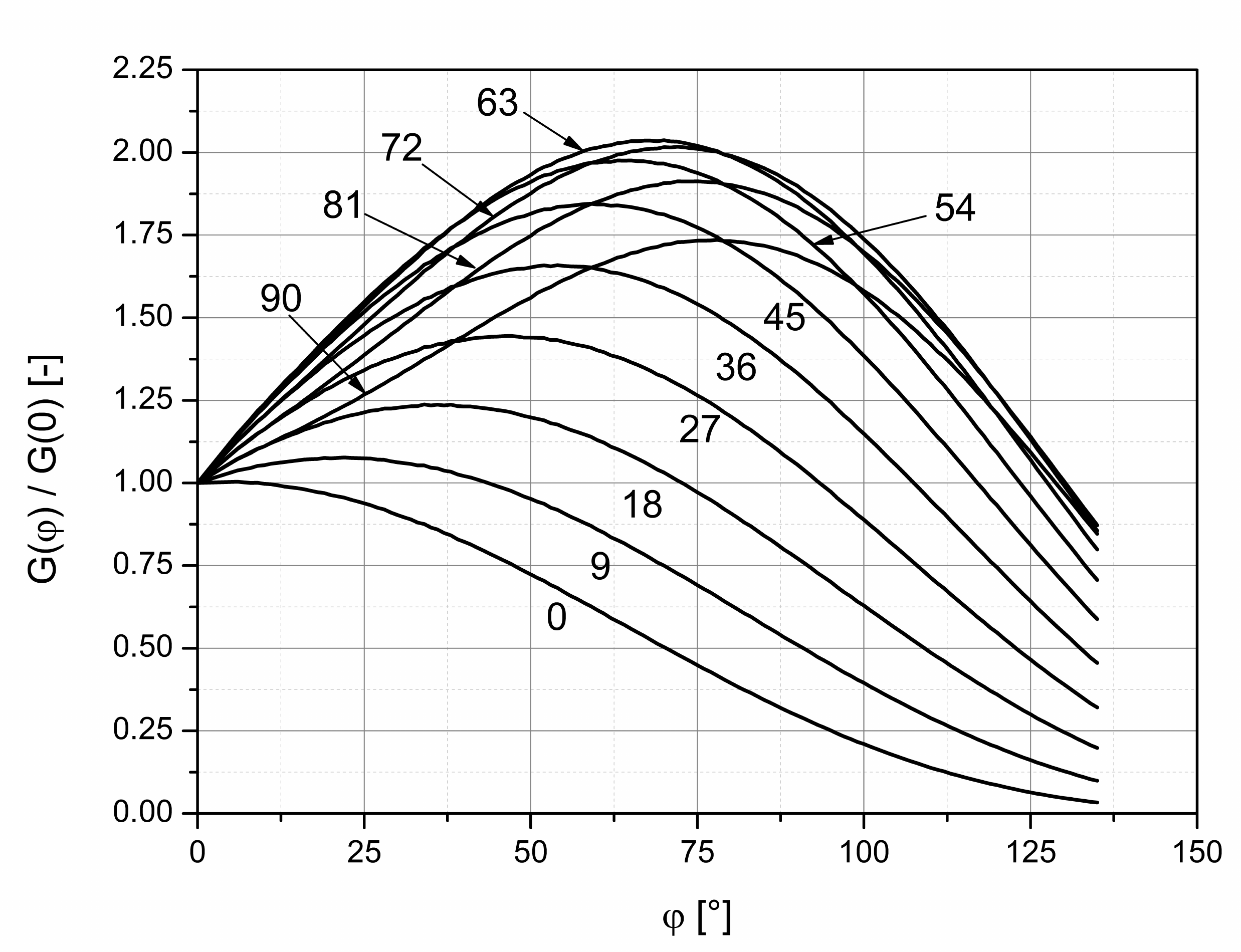} &  
		\includegraphics[width=0.49\textwidth]{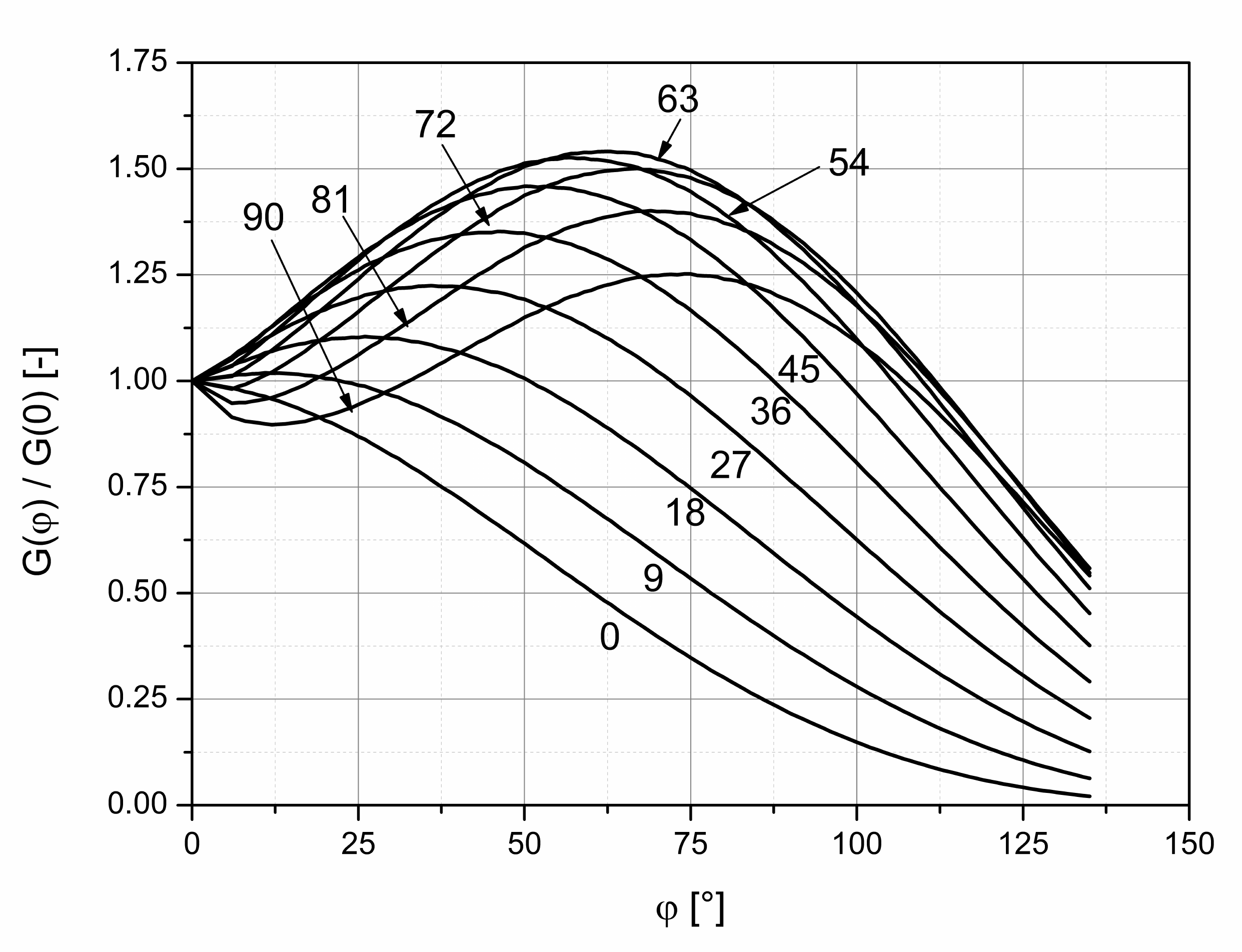} \\
		(i)  &   (ii)  
	\end{tabular}
	\caption{Dependence of the energy release rate ratio $G(\varphi)/G(0)$  as a function of the kink angle $\varphi$ and load combination $\gamma$ for different Dundurs parameter $\alpha$: (i) $\alpha=0.25$, (ii) $\alpha=-0.25$.}
	\label{fig:6}
\end{figure}

\begin{figure}[htp]
	\begin{tabular}{cc}
		\includegraphics[width=0.49\textwidth]{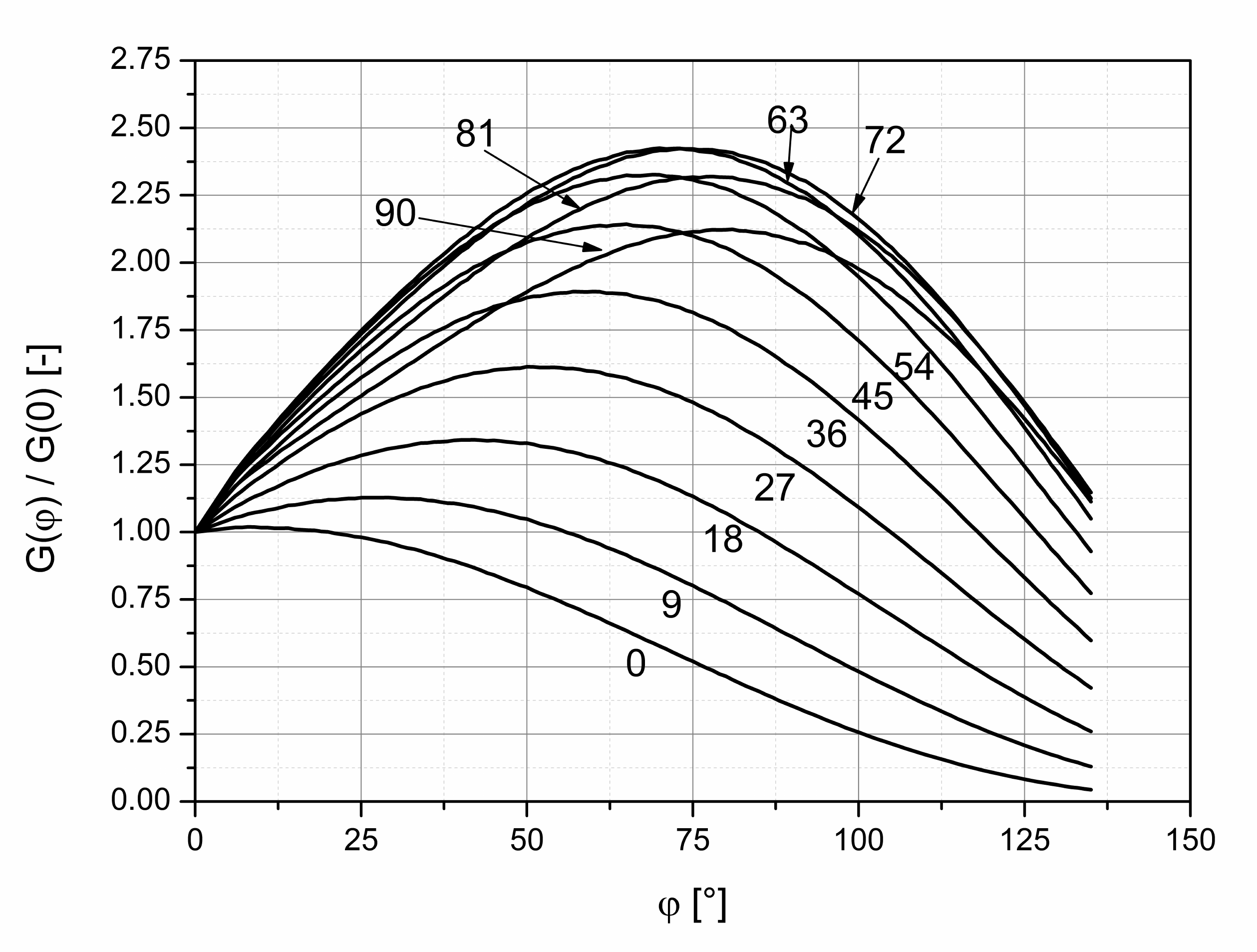} &  
		\includegraphics[width=0.49\textwidth]{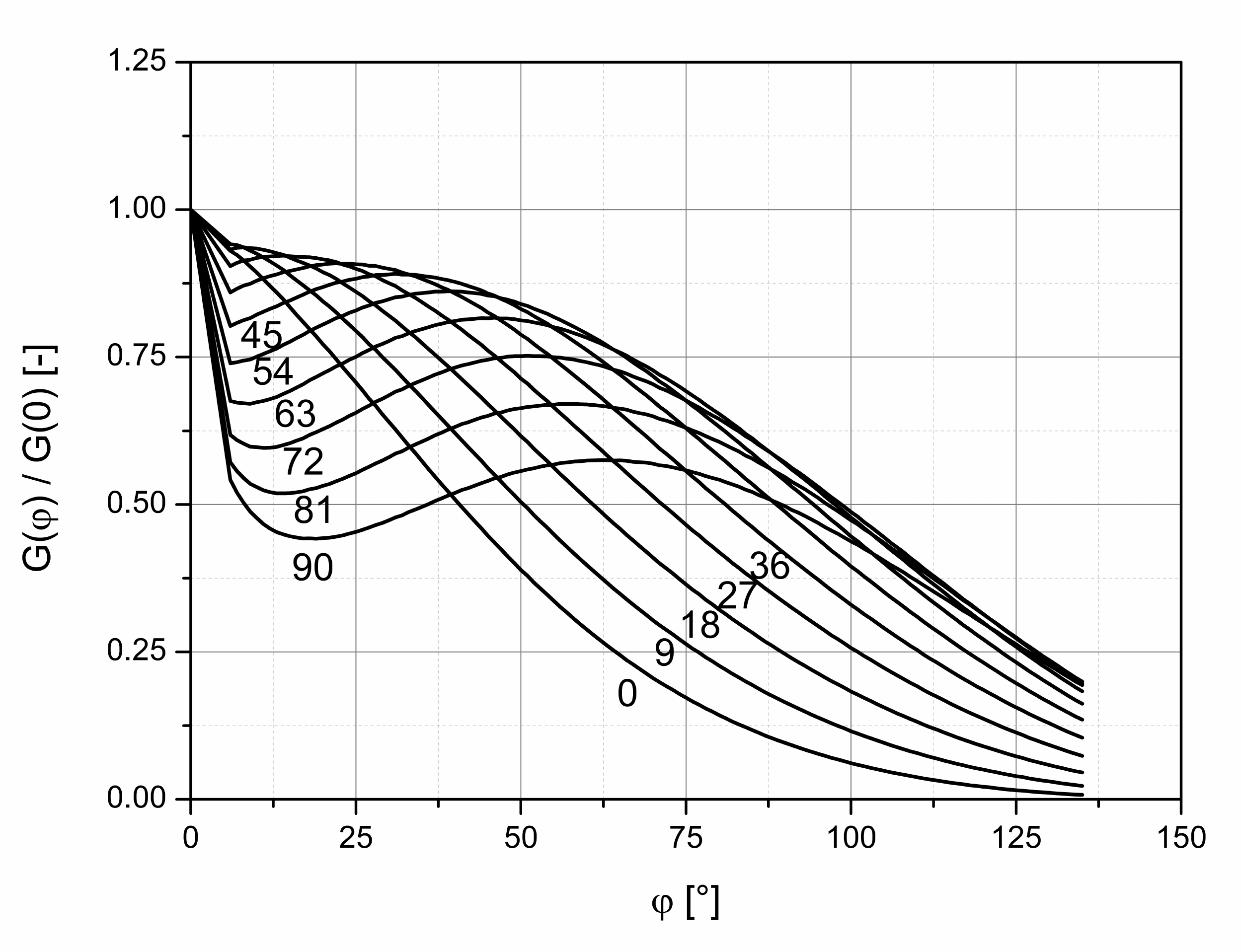} \\
		(i)  &   (ii)  
	\end{tabular}
	\caption{Dependence of the energy release rate ratio $G(\varphi)/G(0)$  as a function of the kink angle $\varphi$ and load combination $\gamma$ for different Dundurs parameter $\alpha$: (i) $\alpha=0.75$, (ii) $\alpha=-0.75$.}
	\label{fig:7}
\end{figure}

\begin{figure}[htp]
	\begin{tabular}{cc}
		\includegraphics[width=0.49\textwidth]{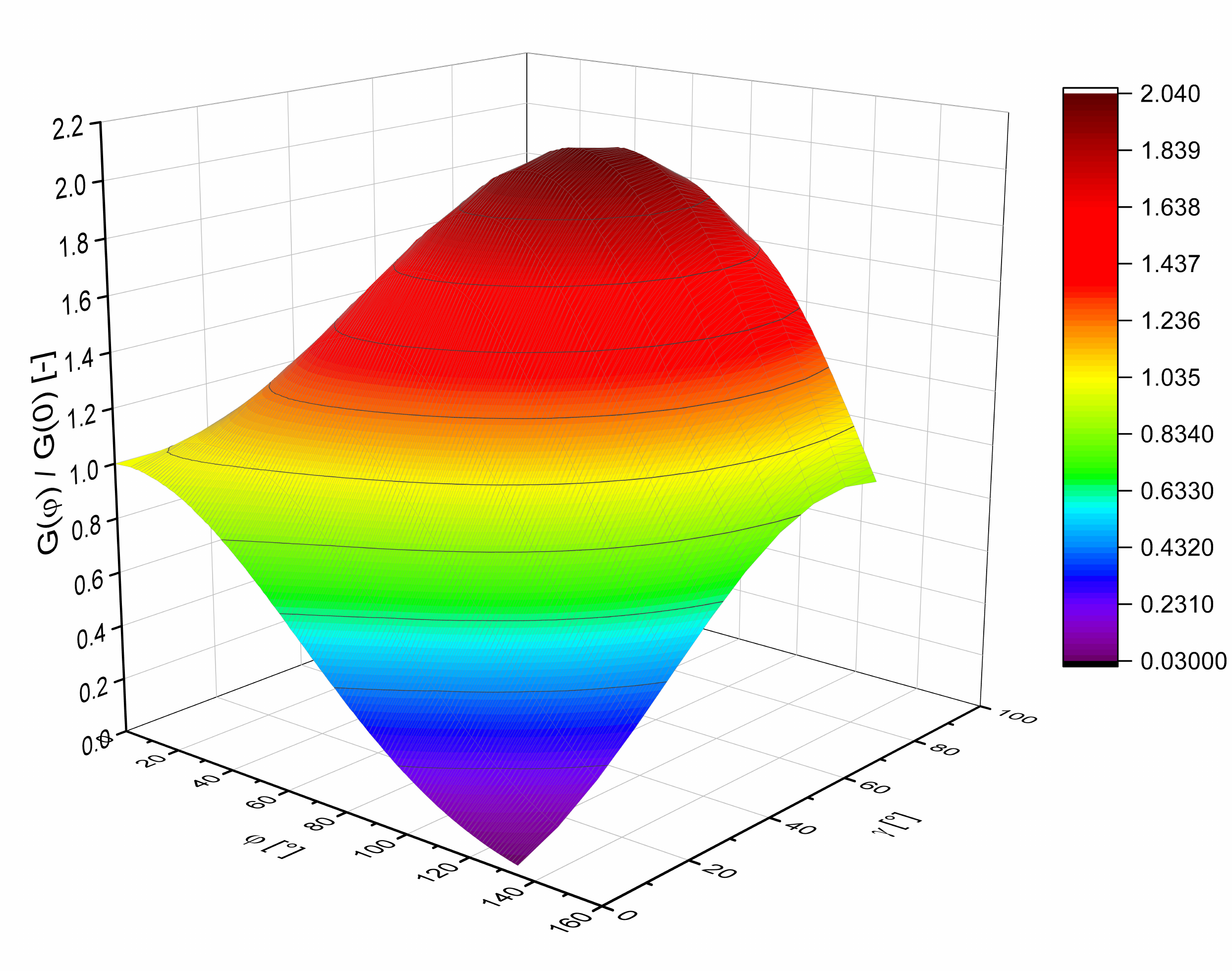} &  
		\includegraphics[width=0.49\textwidth]{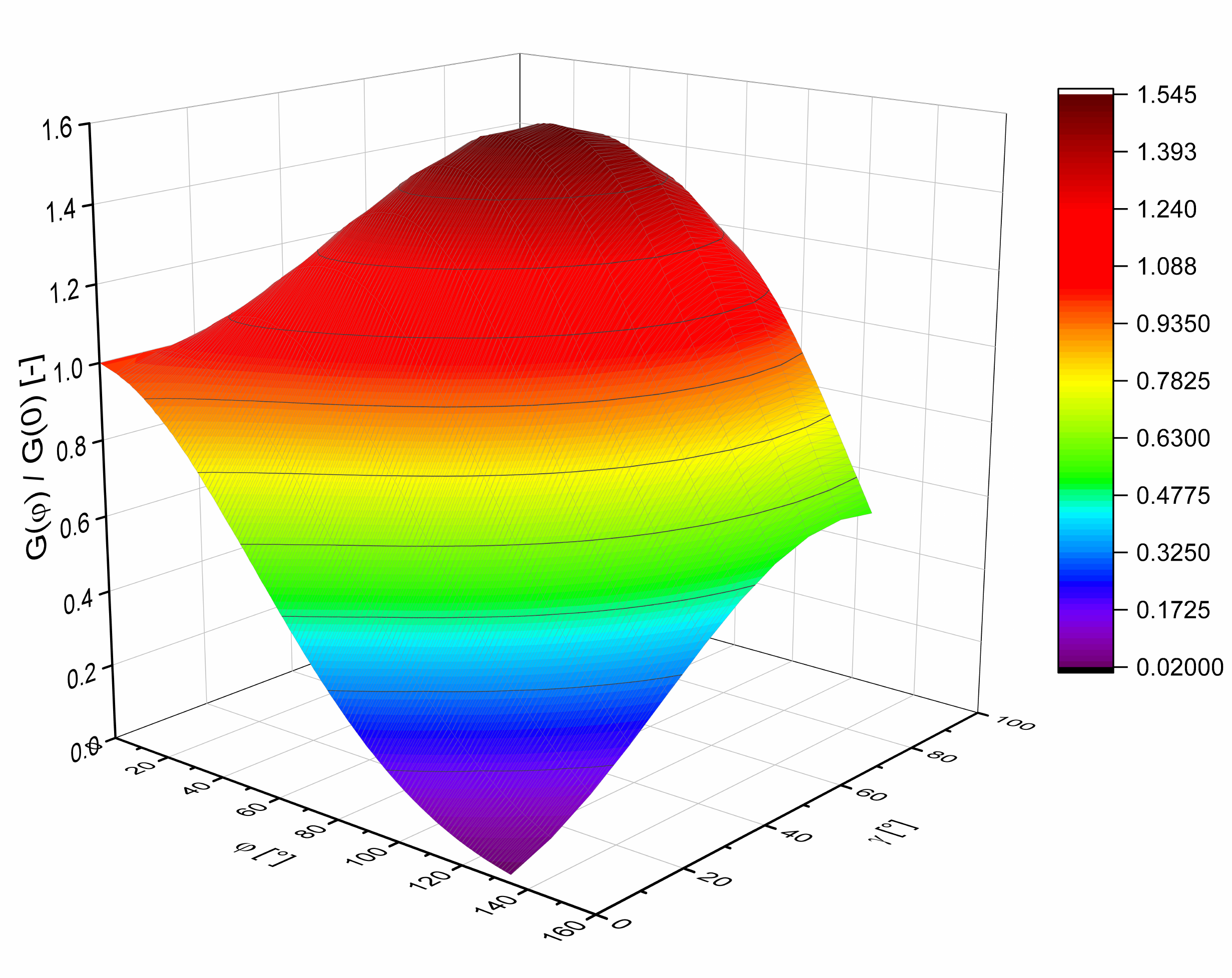} \\
		(i)  &   (ii)  
	\end{tabular}
	\caption{(Color online) 3D-plots of the energy release rate ratio $G(\varphi)/G(0)$  as a function of the kink angle $\varphi$ and load combination $\gamma$ for different Dundurs parameter $\alpha$: (i) $\alpha=0.25$, (ii) $\alpha=-0.25$.}
	\label{fig:6a}
\end{figure}

\begin{figure}[htp]
	\begin{tabular}{cc}
		\includegraphics[width=0.49\textwidth]{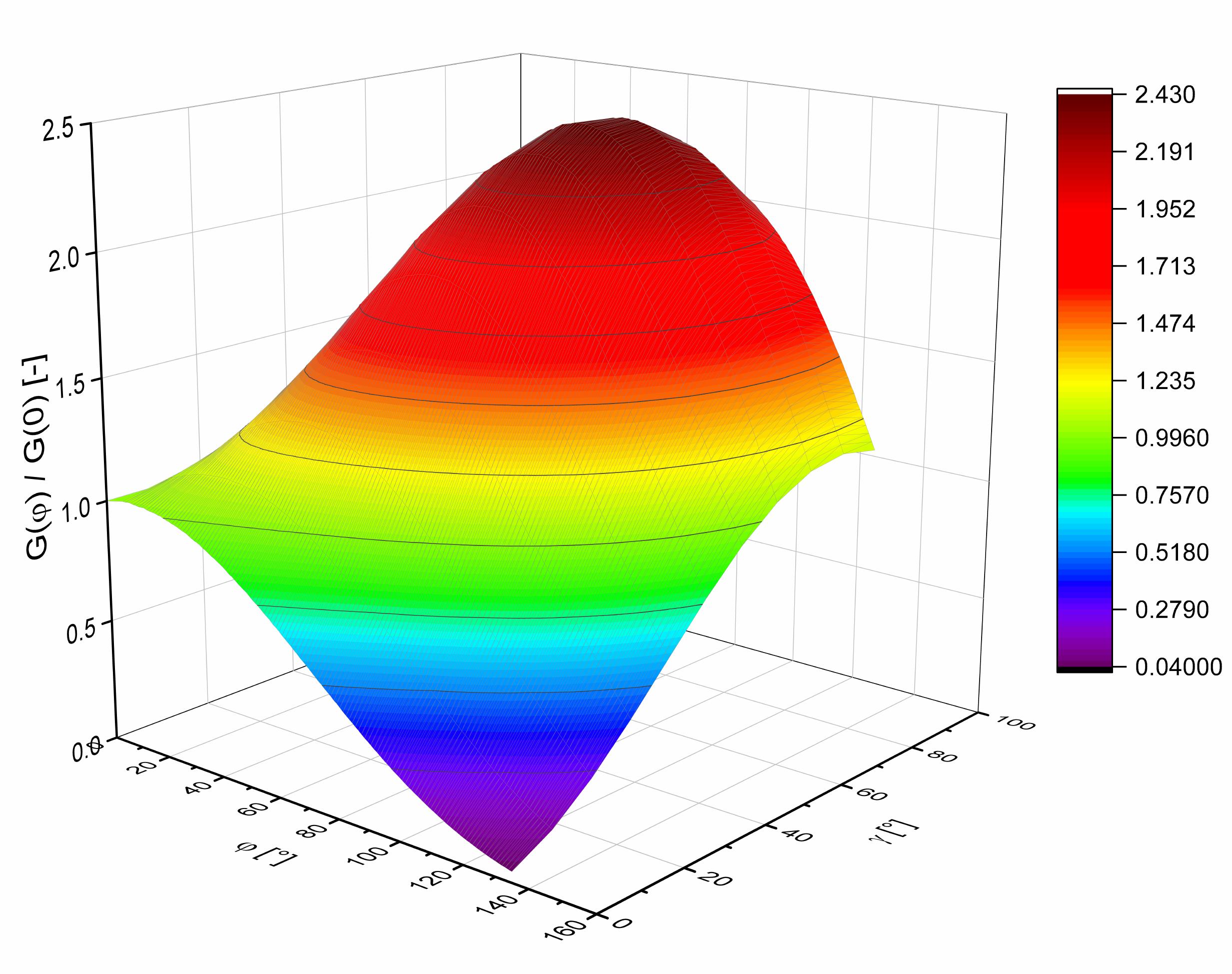} &  
		\includegraphics[width=0.49\textwidth]{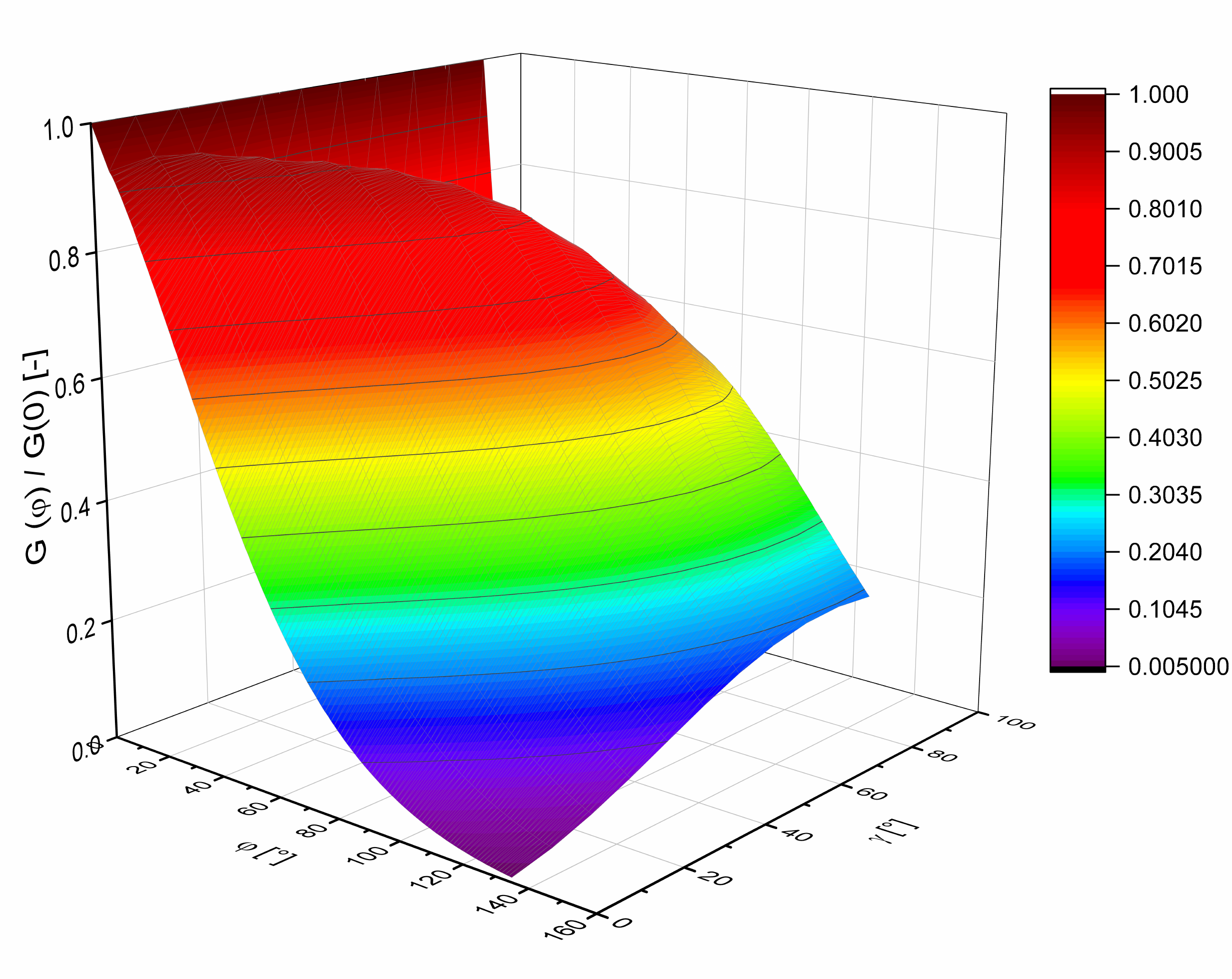} \\
		(i)  &   (ii)  
	\end{tabular}
	\caption{(Color online) 3D-plots of the energy release rate ratio $G(\varphi)/G(0)$  as a function of the kink angle $\varphi$ and load combination $\gamma$ for different Dundurs parameter $\alpha$: (i) $\alpha=0.75$, (ii) $\alpha=-0.75$.}
	\label{fig:7a}
\end{figure}

The finding of coefficients $c$ and $d$ in \eqref{4.2} is based on the numerical solution of the singular integral equation (see \cite{he1989kinking,he1989akinking,noijen2012semi}). Using their results we can find the energy release rate ratio $G(\varphi)/G(0)$, whose plots are shown in Figs.~\ref{fig:6}(i)-(ii) for $\alpha=\pm 0.25$ and \ref{fig:7}(i)-(ii) for $\alpha=\pm 0.75$. The corresponding 3D plots of $G(\varphi)/G(0)$ as function of two variables $\varphi$ and $\gamma$ are shown in Figs.~\ref{fig:6a}(i)-(ii) for $\alpha=\pm 0.25$ and \ref{fig:7a}(i)-(ii) for $\alpha=\pm 0.75$. From these plots, one can see how the variation of the Dundurs mismatch parameter $\alpha$ affects the change in the maximum of $G(\varphi)/G(0)$ with respect to $\varphi$. In particular, the slopes of the curves $G(\varphi)/G(0)$ as a function of $\varphi$ for positive and negative $\alpha$ are different. For large negative $\alpha$, the maximum is less than 1, as can be seen in Fig.~\ref{fig:7}(ii). Using Eq.~\eqref{2.14} in combination with the maximum energy release rate criterion, we can easily predict the direction of crack growth in terms of the Dundurs parameter $\alpha$ and the parameter $\gamma$ characterizing the load combination. 

\begin{figure}[htb]
\centering \includegraphics[width=0.5\textwidth]{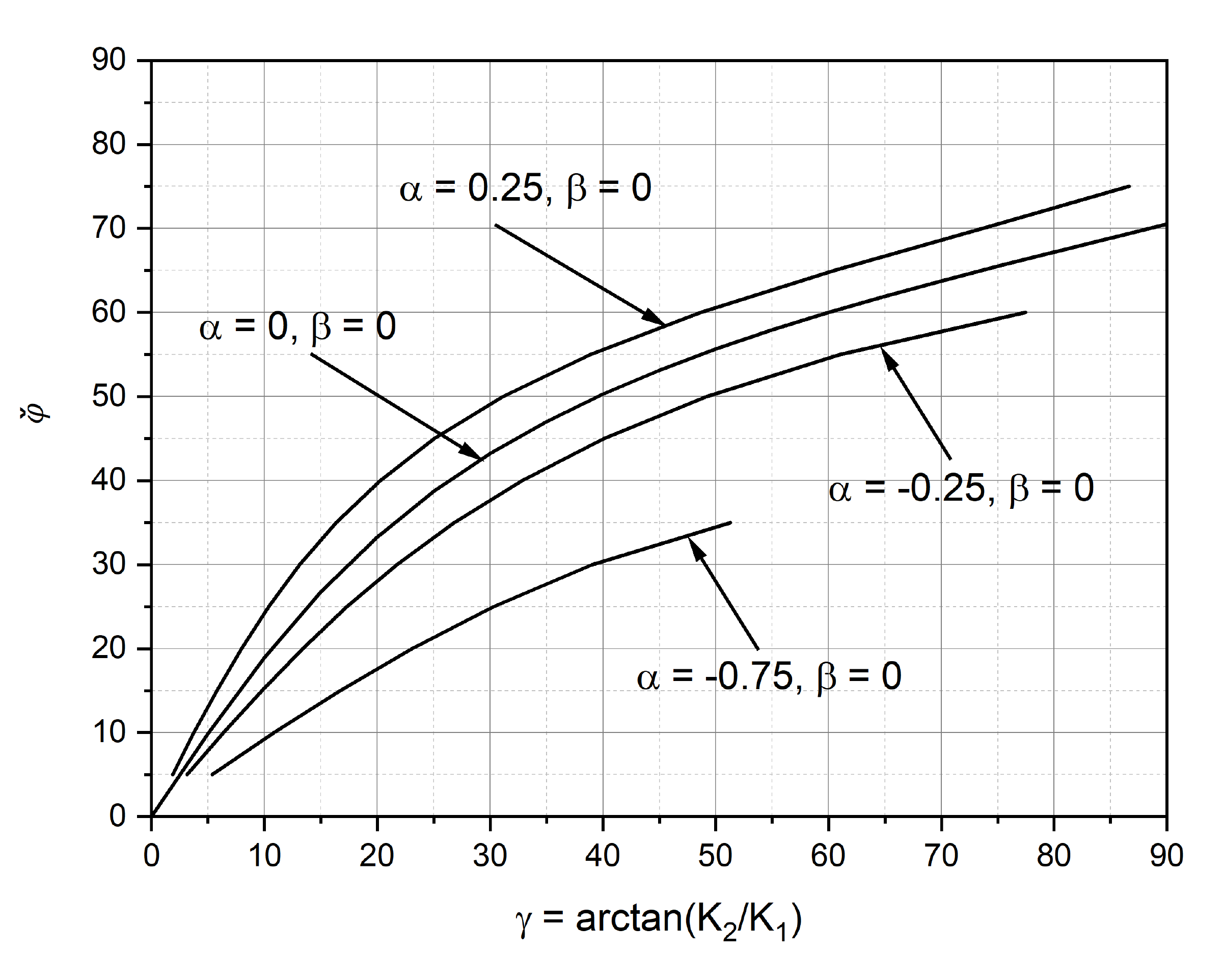} \caption{Kink angle $\check{\varphi}$ as a function of the loading combination $\gamma$ at different $\alpha$.} \label{fig:8}
\end{figure}

The plot of the kink angle $\check{\varphi}$ that maximizes the energy release rate as function of the load combination $\gamma$ at different $\alpha$ is shown in Fig.~\ref{fig:8}. Note that the deviation from that angle determined by the criterion $K_{II}=0$ is small except for large $\gamma$.
 
\section{Conclusion}
The main result of this work is the derivation of the maximum energy reduction rate criterion from the variational principle of fracture mechanics. The derived criterion can be applied both to the interface crack and to the crack whose tip reaches the interface. Although this criterion reduces to the well-known and widely accepted criteria of crack growth formulated by He and Hutchinson \cite{he1989crack}, the derivation from first principles brings two advantages. First, it removes the uncertainties in the choice of criteria mentioned in the Introduction. Second, the variational formulation opens the way for direct numerical methods, e.g., the computationally effective non-remeshing finite elements proposed, for example, in \cite{moes1999finite}, which can be applied to inhomogeneous cracked bodies of arbitrary geometry. 

There are several possible extensions to the variational formulation given in this paper. First, the extended variational problem can be studied for laminated composites with imperfect bonding or thin film adhesives, where the surface energy could depend on the displacement jump. Second, it can be extended to inhomogeneous materials that deform at finite strain, with the goal of applying it to filled elastomers (cf. \cite{stumpf1990variational,le1992singular,le1993singular,el2017prediction} for nonlinear fracture mechanics of homogeneous rubbery materials). Third, an interesting topic closely related to the one considered here is crack nucleation in inhomogeneous solids. To address this issue, thermal fluctuation in the spirit of \cite{berdichevsky2005microcrack} must be considered in addition to the energy of microcracks.

\end{document}